\documentclass[aps, pra, showpacs, superscriptaddress, twocolumn, twocolumn,
preprintnumbers]
{revtex4-1}
\usepackage{graphicx}
\usepackage{calc}
\usepackage{amsmath}
\usepackage{amssymb}

\newcommand{\D}{\mbox{\rm d}}
\newcommand{\Tr}{\mbox{\rm Tr}}
\newcommand{\I}{\mbox{\rm I}}
\renewcommand{\Im}{\mathrm{Im}}
\renewcommand{\Re}{\mathrm{Re}}
\newcommand{\E}{\mathrm{E}}
\newcommand{\Var}{\mathrm{Var}}
\newcommand{\ket}[1]{\left|{#1}\right\rangle}
\newcommand{\bra}[1]{\left\langle{#1}\right|}

\begin{document}
\preprint{PHYSICAL REVIEW A {\bf 85}, 013826 (2012)}
\title{Homodyne detection for atmosphere channels}

\author{A. A. Semenov}
\email[E-mail address: ]{sem@iop.kiev.ua}%
\affiliation{Institut f\"ur Physik, Universit\"{a}t Rostock,
Universit\"{a}tsplatz 3, D-18051 Rostock, Germany}
\affiliation{Institute of Physics, NAS of
Ukraine, Prospect Nauky 46, 03028 Kiev, Ukraine}
\affiliation{Bogolyubov Institute for Theoretical Physics, NAS of Ukraine, Vul.
Metrologichna 14-b, 03680
Kiev, Ukraine}%

\author{F. T\"oppel}
\affiliation{Institut f\"ur Physik, Universit\"{a}t Rostock,
Universit\"{a}tsplatz 3, D-18051 Rostock, Germany}
 \affiliation{Max Planck Institute for the Science of Light, G\"unther-Scharowsky-Stra{\ss}e 1/Bau 24, D-91058 Erlangen, Germany}

\author{D. Yu. Vasylyev}
\affiliation{Bogolyubov Institute for Theoretical Physics, NAS of Ukraine, Vul.
Metrologichna 14-b, 03680
Kiev, Ukraine}%
 \affiliation{Max Planck Institute for the Science of Light, G\"unther-Scharowsky-Stra{\ss}e
 1/Bau 24, D-91058 Erlangen, Germany}

\author{H. V. Gomonay}

\affiliation{Bogolyubov Institute for Theoretical Physics, NAS of Ukraine, Vul.
Metrologichna 14-b, 03680
Kiev, Ukraine}%


\author{W. Vogel}
\affiliation{Institut f\"ur Physik, Universit\"{a}t Rostock,
Universit\"{a}tsplatz 3, D-18051 Rostock, Germany}

\begin{abstract}
We give a systematic theoretical description of homodyne detection
in the case where both the signal and the local oscillator pass
through the turbulent atmosphere. Imperfect knowledge of the
local-oscillator amplitude is effectively included in a noisy
density operator, leading to postprocessing noise. Alternatively,
we propose a technique with monitored transmission coefficient of
the atmosphere, which is free of postprocessing noise.
\end{abstract}


\maketitle

\section{Introduction}
\label{Introduction}

Long-distance quantum communication~\cite{Takesue, Fedrizzi}
necessarily deals with strong unwanted effects of the environment.
In this context, one usually compares two types of channels: optical
fibers and free space. For purposes of quantum optics, it is
important that in fibers~\cite{Mitschke} one usually deals with a
stable intensity attenuation and with strong depolarization
effects. In free-space channels~\cite{Tatarskii, Fante}, the
situation is different: The attenuation randomly fluctuates and the
depolarization effect is negligibly small.

An important method for measuring the quantum-light characteristics
is the technique of balanced homodyne detection~\cite{Yuen, Welsch}.
In this case, the signal field is combined through a 50:50
beam splitter with a strong coherent field, the local oscillator.
The difference of photocounts in two outputs of the beam splitter is
proportional to the field quadrature. By applying this procedure for
different values of the local-oscillator phase, one could get
complete information about the quantum state of the signal.
Particularly, one can reconstruct the density operator in different
representations~\cite{Welsch, VogelK, Smithey}.

The application of homodyne detection for long-distance quantum
communications in free-space channels meets the problem of phase
synchronization between the signal and the local oscillator. A
possible way to overcome this difficulty could be based on the
technique of the optical frequency comb~\cite{OFC}. In this case, the
detected signal will be randomized by the atmosphere with respect to
both the amplitude and the phase~\cite{Semenov1}. A more traditional
way to provide such a synchronization is to derive the signal and
the local oscillator from the same source. However, in this case the
local oscillator will also be affected by the atmospheric
turbulence. At least part of this problem can be resolved by sending
the signal and the local oscillator from the same source in
orthogonally polarized modes~\cite{Elser}. Since the atmospheric
depolarization effects are negligible, the phase synchronization is
not destroyed in such an experiment. On the other hand, in this case
the local-oscillator amplitude randomly fluctuates due to the
atmospheric noise. As a result, the problem is how to connect
the photocount difference with the field quadrature.

In the present paper we consider the situation when the signal and
the local oscillator pass through the turbulent atmosphere in
orthogonally polarized modes. First, we analyze the scheme proposed
in Ref.~\cite{Elser}. In this case, the photocount difference can be
connected with the field quadrature by using a certain reference
value of the local-oscillator amplitude. This is equivalent to the
use of a reference transmission coefficient of the atmosphere, for example
its mean value. This results in a kind of noise, which is related to
the postprocessing procedure. Reconstructed with such a procedure,
the density operator may even fail to satisfy the fundamental
requirement of positive semidefiniteness, which is a serious
disadvantage of this method. To resolve this deficiency, we propose
a procedure with a permanently monitored transmission coefficient.
This renders it possible to recover the true values of the field
quadratures from the measured photocount differences. In
this case, the postprocessing noise disappears.

The paper is organized as follows. In Sec.~\ref{Statistics}, we
derive an expression for the statistics of photocount differences
for the scheme considered in Ref.~\cite{Elser}. This result is used
in Sec.~\ref{Effective}, where the photocount difference is
connected with the field quadrature by using a fixed reference
transmission coefficient. A method based on  monitoring the
transmission coefficient is developed in Sec.~\ref{Monitored}. In
Sec.~\ref{Quadrature}, we derive input-output relations for the
normally ordered covariance matrix and consider the effect of
quadrature squeezing of the light passing through the atmosphere. A
summary and conclusions are given in Sec.~\ref{Conclusions}.

\section{Statistics of photocount differences}
\label{Statistics}

\begin{figure*}[ht!]
\includegraphics{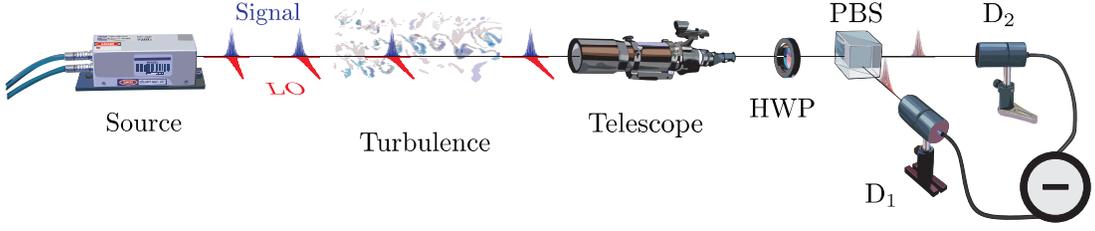}
\caption{\label{fig:scheme} (Color online) Scheme of homodyne
detection for quantum light passing through the turbulent
atmosphere as reported in Ref.~\cite{Elser}. The signal and
the local oscillator are sent in two orthogonally polarized modes.
After passing through the atmospheric channel and collection by a
telescope, homodyne detection is realized with a half-wave plate
$\mathrm{HWP}$, a polarization beam-splitter $\mathrm{PBS}$, and
the detectors $\mathrm{D}_1$ and $\mathrm{D}_2$.}
\end{figure*}

Let us consider an
experimental scenario as implemented in Ref.~\cite{Elser} with the
local oscillator copropagating with the signal field in the same
spatial but different polarization modes; see Fig.~\ref{fig:scheme}.
The half-wave plate $\mathrm{HWP}$ and the polarization
beam-splitter $\mathrm{PBS}$ at the receiver prepare $50{:}50$
field superposition of the signal and the local oscillator as
done in the standard homodyne detection. The detectors
$\mathrm{D}_1$ and $\mathrm{D}_2$ are used for measuring the photocount
difference $\Delta n$, which is used for  further analysis.

The probability distribution of the photocount difference $\Delta
n$ in the Heisenberg picture is given by  (cf.
Ref.~\cite{Grabow})
\begin{equation}
p_{{}_{\Delta n}}=\Tr\left[\hat\rho\, \hat K_{\Delta
n}^\mathrm{noisy}\right],\label{PCD_Op_2}
\end{equation}
where $\hat\rho$ is the input-signal density operator and
$\hat{K}_{\Delta n}^\mathrm{noisy}$ is the noisy positive
operator-valued measure (POVM) of photocount differences. The
latter,
 \begin{equation}
  \hat{K}_{\Delta n}^\mathrm{noisy}=\bra{re^{i\varphi}
}\bra{0}\sum\limits_{n=\Delta n}^{+\infty}
\hat{\Pi}_{n}\otimes\hat{\Pi}_{n-\Delta
n}\ket{0}\ket{re^{i\varphi}},\label{POVM_DN_OP}
 \end{equation}
is expressed in terms of the POVM of photocounts,
\begin{equation}
 \hat{\Pi}_{n_i}=
:\frac{\left(\eta\,\hat{b}_{i}^{\dag}\hat{b}_{i}\,+\bar{N}_\mathrm{nc}\right)^{n_i}}{
n_i! }
\exp\left(-\eta\,\hat{b}_{i}^{\dag}\hat{b}_{i}-\bar{N}_\mathrm{nc}
\right):,
\label{POVM}
\end{equation}
where $:\cdots :$ denotes the normal ordering prescription. The
coherent-state vector $\ket{re^{i\varphi}}$ represents the local
oscillator of amplitude $r$ and phase $\varphi$. The  vacuum-state
vector $\ket{0}$ includes all modes of the environment. Moreover,
$\eta$ is the detection efficiency, and $\bar{N}_\mathrm{nc}$ is the
mean value of noise counts caused by stray light as well as dark
counts; see Ref.~\cite{Semenov3}. The annihilation and creation
operators $\hat{b}_{i}$ and $\hat{b}_{i}^{\dag}$, respectively,
represent the light modes at the $i\mathrm{th}$ output of the
polarization beam splitter $\mathrm{PBS}$. This beam splitter and
the half wave-plate $\mathrm{HWP}$ can be described by the
input-output relations,
\begin{eqnarray}
&\hat{b}_{1}=\frac{1}{\sqrt{2}}\left(\hat{b}_\mathrm{s}+
\hat{b}_\mathrm{lo}\right),
\label{IOR1}\\
&\hat{b}_{2}=\frac{1}{\sqrt{2}}\left(-\hat{b}_\mathrm{s}+
\hat{b}_\mathrm{lo}
\right),\label{IOR2}
\end{eqnarray}
where $\hat{b}_\mathrm{s}$ and $\hat{b}_\mathrm{lo}$ are annihilation
operators of the signal and the local oscillator at the input of the
beam splitter.

Next, we have to include in the consideration the effect of the
atmosphere. It can be performed using the approach of
fluctuating-loss channels~\cite{Semenov1, Semenov2, Vasylyev, Dong,
Shapiro}. Let $\hat{a}_\mathrm{s}$ and $\hat{a}_\mathrm{lo}$ be the
annihilation operators of the signal and the local oscillator,
respectively, at the sender. The corresponding input-output relation
for light passing through the atmosphere reads as
\begin{eqnarray}
&\hat{b}_\mathrm{s}=T\hat{a}_\mathrm{s}+\sqrt{1-T^2}\hat{c}_\mathrm{s} ,
\label{IORAs}\\
&\hat{b}_\mathrm{lo}=T\hat{a}_\mathrm{lo}+\sqrt{1-T^2}\hat{c}_\mathrm{lo},
\label{IORAlo}
\end{eqnarray}
where $\hat{c}_\mathrm{s}$ and $\hat{c}_\mathrm{lo}$ are operators
of the environment modes being in the vacuum state and $T$ is the
atmospheric transmission coefficient. The following properties of
relations (\ref{IORAs}) and (\ref{IORAlo}) are important. First, the
transmission coefficient $T$ is a random variable. Second, since the
depolarization effect of the atmosphere is negligible, the
transmission coefficients in both polarization modes are perfectly
correlated and equal $T$. Third, in the considered case the absence
of the depolarization means the absence of dephasing, and hence $T$ can
be considered as a real random variable. Finally, the commutation
rules require that $T\in [0,1]$.

The above treatment can be easily used in Eq.~(\ref{PCD_Op_2}) with
the Glauber-Sudarshan $P$ representation~\cite{GlauberSudarshan} for
the signal density operator,
\begin{equation}
 \hat{\rho}=\int\limits_{-\infty}^{+\infty}\D^2\alpha \ket{\alpha}P\!\left(\alpha\right)\bra{\alpha},\label{GSF}
\end{equation}
where $P\!\left(\alpha\right)$ is the Glauber-Sudarshan $P$ function
of the input signal field (at the sender) and $\ket{\alpha}$ is a
coherent-state vector. Substituting
Eqs.~(\ref{POVM_DN_OP})--(\ref{GSF}) into Eq.~(\ref{PCD_Op_2}) and
taking into account that $T$ is a random variable, one gets
\begin{equation}
 p_{{}_{\Delta n}}=\int\limits_{-\infty}^{+\infty}\D^2\alpha P\!\left(\alpha\right)
K_{\Delta n}^\mathrm{noisy}\left(\alpha\right),\label{PCD_GS}
\end{equation}
where
\begin{widetext}
\begin{equation}
 K_{\Delta n}^\mathrm{noisy}\left(\alpha\right)=\int\limits_{0}^{1}\D T
\mathcal{P}\!\left(T\right) \left(\frac{\eta\theta_1+2\bar
N_\mathrm{nc}}{\eta\theta_2+2\bar N_\mathrm{nc})}\right)^{\Delta
n/2} \I_{\Delta n}\left[\sqrt{(\eta\theta_1+2\bar
N_\mathrm{nc})(\eta\theta_2+2\bar
N_\mathrm{nc})}\right]\exp{\left[-\eta(T^2r^2+T^2|\alpha|^2)-2\bar
N_\mathrm{nc}\right]}\label{POVM_Q1}
\end{equation}
\end{widetext}
is the Husimi-Kano $Q$ symbol~\cite{HusimiKano} of the POVM of
photocount differences, where
\begin{equation}
  \theta_{1,2}=T^2\left(r^2+|\alpha|^2\pm
  2r\,\Re\!\left[\alpha
e^{-i\varphi}\right]\right),\label{def_theta}
\end{equation}
$\mathcal{P}\!\left(T\right)$ is the probability distribution of the
transmission coefficient (PDTC) of the atmospheric channel and
$\I_{\Delta n}$ is the modified Bessel function. For simplicity,
further we  refer to the $Q$ symbol of the POVM as the POVM.

For the purposes of balanced homodyne detection one usually uses a
local oscillator, which is strong compared to the signal. After
transmission through the atmosphere, the intensity of the signal is
still small compared to the local oscillator,
$T^2\left|\alpha\right|^2\ll T^2 r^2$. The contribution of the
latter can be comparable with the noise counts, $T^2 r^2 \sim \bar
N_\mathrm{nc}$. Following the argumentation of Ref.~\cite{Grabow},
one can approximate Eq. (\ref{POVM_Q1}) by
\begin{eqnarray}
 K_{\Delta n}^\mathrm{noisy}\left(\alpha\right)=\int\limits_{0}^{1}\D T
\mathcal{P}\!\left(T\right) \frac{1}{\sqrt{2\pi(\eta T^2 r^2+2\bar
N_\mathrm{nc})}}\nonumber\\
\times \exp\left[-\frac{(\Delta
n-2\eta T^2 r\, \Re\!\left[\alpha
e^{-i\varphi}\right])^2}{2(\eta T^2 r^2+2\bar
N_\mathrm{nc})}\right].\label{POVM_Q2}
\end{eqnarray}
Equations~(\ref{PCD_GS}) and (\ref{POVM_Q2}) can be directly used
for evaluating the statistics of photocount differences when both
the signal and the local oscillator pass through the turbulent
atmosphere.

\section{Postprocessing noise}
\label{Effective}

The next problem is to connect the photocount differences
$\Delta\hat{n}$ with the field quadrature,
\begin{equation}
 \hat{x}\left(\varphi\right)=\frac{1}{\sqrt{2}}\left[\hat{a}^\dag
e^{i\varphi}+\hat{a} e^{-i\varphi}\right]. \label{Eq:Quadrature}
\end{equation}
The corresponding relation is given by  (see, e.g.,
Ref.~\cite{Welsch}),
\begin{equation}
 \hat{x}\left(\varphi\right){=}\frac{\Delta\hat{n}}
{r_\mathrm{out} \sqrt{2}},\label{X-DN-NN}
\end{equation}
where $r_\mathrm{out}$ is the local-oscillator amplitude scaled by
atmosphere losses and detection efficiency. Its real value is
\begin{equation}
r_\mathrm{out}=T \sqrt{\eta} r. \label{r_out}
\end{equation}
However, since in the considered experiment we do not have any
information about the current value of the fluctuating transmission
coefficient $T$, we can use a certain reference value
$T_\mathrm{ref}$, and set
\begin{equation}
r_\mathrm{out} = T_\mathrm{ref} \sqrt{\eta} r.\label{r_out_ref}
\end{equation}
Because the real value of $r_\mathrm{out}$ given by
Eq.~(\ref{r_out}) randomly changes in the atmosphere and deviates
from its reference value~(\ref{r_out_ref}), the obtained quadrature
value suffers from a kind of noise, which we  refer to as
post-processing noise. For this reason, the reconstructed density
operator and any characteristics obtained from the approximate
quadrature values are, in fact, contaminated by the corresponding
noise effects.

Let us consider in more detail the effects of the postprocessing noise.
Based on Eqs.~(\ref{PCD_GS}), (\ref{POVM_Q2}), (\ref{X-DN-NN}), and
(\ref{r_out_ref}), the quadrature distribution in the considered
case is given by
 \begin{equation}
 p\!\left(x;\varphi\right)=\int\limits_{-\infty}^{+\infty}
 \D^2\alpha\, P\!\left(\alpha\right)
K_\mathrm{noisy}\!\left[x\!\left(\varphi\right);\alpha\right],
\label{PX_N1}
\end{equation}
where
\begin{eqnarray}
 K_\mathrm{noisy}\!\left[x\!\left(\varphi\right);\alpha\right]=
 \int\limits_{0}^{1}\D T\mathcal{P}\!\left(T\right)
\frac{1}{\sqrt{\pi\left(\frac{T^2}{T_\mathrm{ref}^{2}}+ \frac{2\bar
N_\mathrm{nc}}{\eta r^2T_\mathrm{ref}^{2}}\right)}}
\nonumber\\
\times
\exp\left[-\frac{\left(x-\sqrt{2\,\eta}\,\frac{T^2}{T_\mathrm{ref}}\,
\Re\!\left[\alpha
e^{-i\varphi}\right]\right)^2}{\frac{T^2}{T_\mathrm{ref}^{2}}+
\frac{2\bar N_\mathrm{nc}}{\eta
r^2T_\mathrm{ref}^{2}}}\right]\label{POVM_QN2}
\end{eqnarray}
is the resulting noisy quadrature POVM. Alternatively,
Eq.~(\ref{PX_N1}) can be rewritten in the Schr\"odinger picture as
\begin{equation}
 p\!\left(x;\varphi\right)=\int\limits_{-\infty}^{+\infty}\D^2\alpha\,
P_\mathrm{noisy}\!\left(\alpha\right)
K\!\left[x\!\left(\varphi\right);\alpha\right],\label{PX_GS}
\end{equation}
where
\begin{equation}
 K\!\left[x\!\left(\varphi\right);\alpha\right]=\frac{1}{\sqrt{\pi}}
\exp\left[-\left(x-\sqrt{2}\,\Re\!\left[\alpha e^{-i\varphi}\right]\right)^2\right]
\label{POVM_X_GS}
\end{equation}
is the noiseless quadrature POVM,
\begin{eqnarray}
&&P_\mathrm{noisy}\!\left(\alpha\right)=\int\limits_{0}^{1}\D
T\mathcal{P}\!\left(T\right)\frac{T_\mathrm{ref}^{2}}{T^4\eta}\label{P_NonMonit}
\\
&&\times\exp\left[\left(\frac{T^2-T_\mathrm{ref}^{2}}{8T_\mathrm{ref}^{2}}+\frac{\bar
N_\mathrm{nc}}{4 r^2
T_\mathrm{ref}^{2}\eta}\right)\Delta_\alpha\right]
P\!\left(\frac{T_\mathrm{ref}}{T^2\sqrt{\eta}}\,\alpha\right)\nonumber
\end{eqnarray}
is the noisy $P$ function of the detected signal, and
$\Delta_\alpha{=}\frac{\partial^2}{\partial^2\Re\alpha}+
\frac{\partial^2}{\partial^2\Im\alpha}$ is the Laplace operator in
phase space. Equation~(\ref{P_NonMonit}) can be considered as the
quantum-state input-output relation, where the noisy density
operator, represented by the $P$ function, is affected by (i)
fluctuating losses due to the signal transmission through the
atmosphere, (ii) detection losses and noise counts, (iii)
postprocessing noise caused by imperfect knowledge of the
transmission coefficient. Any reconstruction of the density operator
using homodyne-detection data, obtained from Eq.~(\ref{X-DN-NN})
together with the approximation~(\ref{r_out_ref}), will yield the
noisy quantum state~(\ref{P_NonMonit}). Similarly, any
characteristics obtained from such quadratures also correspond to
the noisy density operator.

As  follows from Eq.~(\ref{P_NonMonit}), the contribution of noise
counts in the considered scheme can always be made negligible by
choosing a sufficiently strong local oscillator. In this case, the
quantum-state input-output relation reduces to
\begin{eqnarray}
P_\mathrm{noisy}\!\left(\alpha\right)&&=\int\limits_{0}^{1}\D
T\,\mathcal{P}\!\left(T\right)\frac{T_\mathrm{ref}^{2}}{T^4\eta}\label{P_NonMonitN0}\\
&&\times\!\exp\!\left[\frac{T^2-T_\mathrm{ref}^{2}}{8T_\mathrm{ref}^{2}}\Delta_\alpha\right]
P\!\left(\frac{T_\mathrm{ref}}{T^2\sqrt{\eta}}\,\alpha\right).\nonumber
\end{eqnarray}
This equation can be interpreted as the input-output relation of a
fluctuating-loss channel  (cf. Ref.~\cite{Semenov1}) with the
effective transmission coefficient
\begin{equation}
T_\mathrm{eff}=\frac{T^2}{T_\mathrm{ref}}.\label{TrCoeffEff}
\end{equation}
Additionally, the measurement procedure suffers from a kind of
effective noise counts whose mean value is
\begin{equation}
\bar N_\mathrm{eff}\sim
\frac{T^2-T_\mathrm{ref}^{2}}{8T_\mathrm{ref}^{2}}.\label{NcountsEff}
\end{equation}
However, the upper bound of $T_\mathrm{eff}$ is not restricted
anymore by the value $1$. Similarly, $\bar N_\mathrm{eff}$ may take
negative values. As a result, the noisy density operator obtained by
using $P_\mathrm{noisy}\!\left(\alpha\right)$ in Eq.~(\ref{GSF}) may
fail to obey the requirement of positive semidefiniteness. Such an
unusual result simply reflects quantum physical inconsistencies of
the method of data post processing under consideration.

We consider an illustration of this fact for a single-photon-added
thermal state (SPATS). This state is obtained from the single-mode thermal
state with the mean photon number $\bar{n}_\mathrm{th}$ by adding a
photon by using parametric down-conversion. The corresponding $P$
function,
\begin{equation}
P\left(\alpha\right)=\frac{1}{\pi
\bar{n}_\mathrm{th}^3}\left[\left(1+\bar{n}_\mathrm{th}
\right)\left|\alpha\right|^2-\bar{n}_\mathrm{th}
\right]e^{-\frac{\left|\alpha\right|^2}{\bar{n}_\mathrm{th}}},
\label{SPATS_0}
\end{equation}
is regular, which allows its experimental
reconstruction~\cite{Kiesel}. We also assume that fluctuating losses
are caused by beam wandering only; see the appendix~\ref{AppendixA}.
This results in log-negative Weibull distribution for the PDTC  [cf.
Eq.~(\ref{Weibull})].

\begin{figure}[ht!]
\mbox{\includegraphics{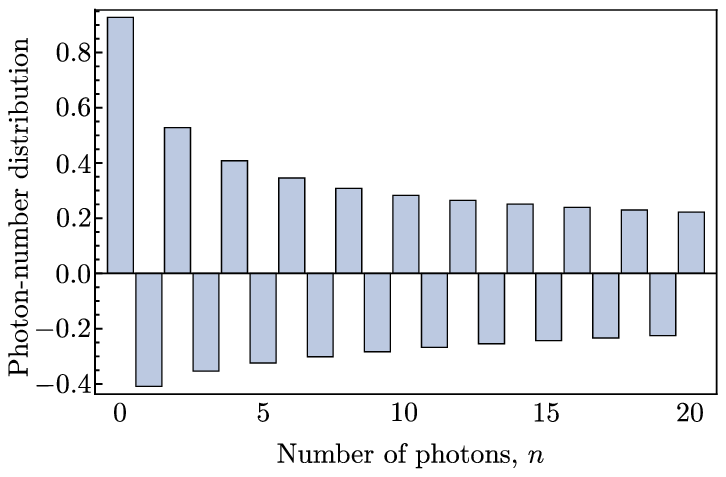}\bf{(a)}}\\
\mbox{\includegraphics{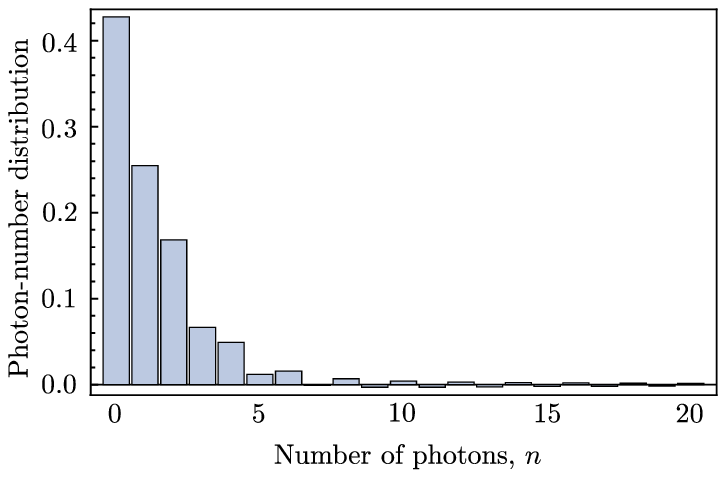}\bf{(b)}} \caption{\label{fig:SPATS_N}
(Color online) Photon-number distribution (diagonal elements of the
effective density operator) for the SPATS with
$n_\mathrm{th}{=}1.11$ (such as realized in Ref.~\cite{Kiesel})
disturbed by beam wandering for the scheme with unmonitored
transmission coefficient. In the corresponding turbulence model,
cf.~the appendix~\ref{AppendixA}, the beam-spot radius, $W$, on the
receiver aperture of radius $a$ is $W{=}0.9 a$,  and the standard
deviations of beam deflection are (a) $\sigma{=}a$ (strong turbulence)
and (b) $\sigma{=}0.5 a$ (weak turbulence). The corresponding
reference transmission coefficient is
$T_\mathrm{ref}{=}\sqrt{\left\langle T^2 \right\rangle}\approx0.586$
and $T_\mathrm{ref}{=}\sqrt{\left\langle T^2
\right\rangle}\approx0.822$, respectively. For both cases we use the
detection efficiency $\eta{=}0.5$. Negative probabilities
demonstrate fake effects caused by imperfect postprocessing.}
\end{figure}

For a consistent positive-definite density operator, the diagonal
elements in the Fock-number basis, that is, photon-number distribution,
should be always nonnegative. This distribution can be
reconstructed from the homodyne-detection data; see
Refs.~\cite{Welsch} and \cite{Munroe}. In the strong-turbulence regime,
the postprocessing noise may result in negative values
of the photon-number distribution; see Fig.~\ref{fig:SPATS_N}(a).
However, such fake effects can be substantially reduced in the case of weak
turbulence; see Fig.~\ref{fig:SPATS_N}(b).

\section{Monitored turbulence}
\label{Monitored}

\begin{figure*}[th!]
\includegraphics{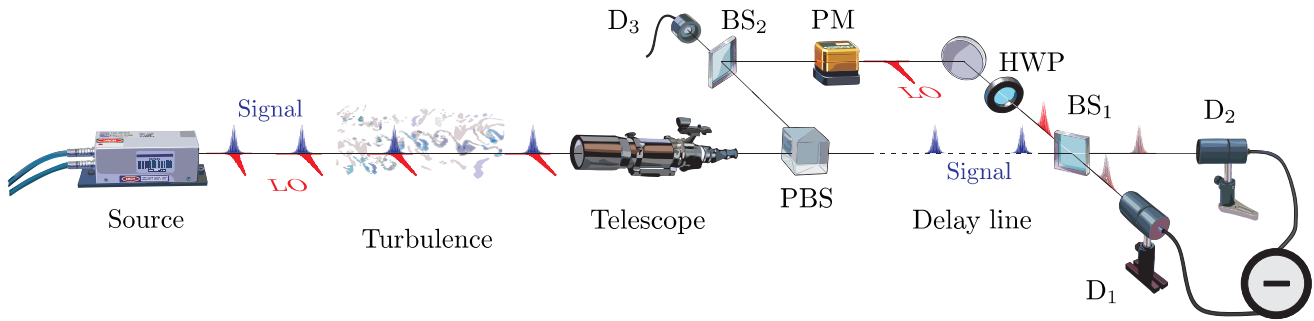}
\caption{\label{fig:scheme_mod} (Color online)
Homodyne detection of quantum light passing through the turbulent
atmosphere with monitoring the transmission coefficient. The signal
and the local oscillator are sent in two orthogonally polarized
modes. After passing through the atmospheric channel and collection
by a telescope, the signal and the local oscillator are split by the
polarization beam-splitter $\mathrm{PBS}$. A part of the local
oscillator transmitted through the beam-splitter $\mathrm{BS}_2$ is
used for monitoring the transmission coefficient with the detector
$\mathrm{D}_3$. Another part, after setting the needed phase by the
phase modulator $\mathrm{PM}$ and conversion of the polarization
direction by the half-wave plate $\mathrm{HWP}$, is combined with
the signal on a standard homodyne detector, which consists of the
$50:50$ beam splitter $\mathrm{BS}_1$ and two photodetectors
$\mathrm{D}_1$ and $\mathrm{D}_2$.}
\end{figure*}

In order to exclude unwanted and fake effects of the postprocessing
noise, we propose to modify the scheme in Fig.~\ref{fig:scheme}. For
this purpose, the signal and the local oscillator can be split, and a
part of the local oscillator can be used for monitoring the current
value of the transmission coefficient; see
Fig.~\ref{fig:scheme_mod}. In the most general case the monitoring
will also be affected by different kinds of noise, for example by
the shot noise of the detector $\mathrm{D}_3$. In terms of the
previous section, this means that the value $T_\mathrm{ref}$ can now
be replaced by the measured transmission coefficient
$T_\mathrm{meas}$, which randomly fluctuates and correlates with
fluctuating values of $T$. The corresponding noisy density operator
is obtained similarly to Eq.~(\ref{P_NonMonit}). However, the PDTC
must now be replaced with the joint PDTC
$\mathcal{P}\!\left(T,T_\mathrm{meas}\right)$ and integrated over
both variables $T$ and $T_\mathrm{meas}$. The joint PDTC can be
given as
\begin{equation}
\mathcal{P}\!\left(T,T_\mathrm{meas}\right)=\mathcal{P}\!\left(T\right)\,
\mathcal{P}\!\left(T_\mathrm{meas}|T\right), \label{cond_PDTC}
\end{equation}
where $\mathcal{P}\!\left(T_\mathrm{meas}|T\right)$ is the
probability distribution of the measured transmission coefficient
under the condition that its real value is $T$. This implies that
the input-output relation for the considered experimental scheme
reads as
\begin{widetext}
\begin{equation}
P_\mathrm{noisy}\!\left(\alpha\right)=\int\limits_{0}^{1}\D
T\int\limits_{0}^{1}\D
T_\mathrm{meas}\,\mathcal{P}\!\left(T\right)\,\mathcal{P}\!
\left(T_\mathrm{meas}|T\right) \frac{T_\mathrm{meas}^{2}}{T^4\eta}
\exp\left[\left(\frac{T^2-T_\mathrm{meas}^{2}}{8T_\mathrm{meas}^{2}}+\frac{\bar
N_\mathrm{nc}}{4 |R_2|^2 r^2
T_\mathrm{meas}^{2}\eta}\right)\Delta_\alpha\right]
P\!\left(\frac{T_\mathrm{meas}}{T^2\sqrt{\eta}}\,\alpha\right),
\label{P_Monit_Noisy}
\end{equation}
\end{widetext}
where $R_2$ is the reflection coefficient of the
beam splitter $\mathrm{BS_2}$.

Let us consider the situation when the noise of monitoring is caused
by the shot-noise of the detector $\mathrm{D}_3$. In this case, the
detected number of photocounts $n_3$ is related to the measured
transmission coefficient $T_\mathrm{meas}$ as
$n_3=r^2\eta_{{}_3}|T_2|^2T_\mathrm{meas}^2+\bar{N}_3$. Here
$\eta_{{}_3}$ and $\bar{N}_3$ are the efficiency and the mean number
of noise counts of detector $\mathrm{D}_3$, respectively; $T_2$ is
the transmission coefficient of the beam splitter $\mathrm{BS}_2$.
The measured transmission coefficient $T_\mathrm{meas}$ is thus
obtained from the number of photocounts $n_3$ via
\begin{equation}
T_\mathrm{meas}^2\!\left(n_{{}_3}\right)=\frac{1}{r^2|T_2|^2}
\frac{n_{{}_3}-\bar{N}_3}{\eta_{{}_3}}.\label{Tr_n3}
\end{equation}
Since detector $\mathrm{D}_3$ records a fraction of the local
oscillator, $n_{{}_3}$ obeys the Poissonian statistics
\begin{equation}
p\left(n_{{}_3}|T\right)=
\frac{\left(\eta_{{}_3}r^2|T_2|^2T^2+\bar{N}_3\right)^{n_{{}_3}}}
{n_{{}_3}!}e^{-\eta_{{}_3}r^2|T_2|^2T^2-\bar{N}_3}.\label{Poisson}
\end{equation}
This is the probability to get the photocount number $n_3$,
conditioned on the value $T$ of the transmission coefficient.

The conditional PDTC $\mathcal{P}\!\left(T_\mathrm{meas}|T\right)$
is obtained, by transforming random variables $n_3 \to
T_\mathrm{meas}$ and using Eq.~(\ref{Tr_n3}), in the form
\begin{equation}
\mathcal{P}\!\left(T_\mathrm{meas}|T\right)=
\sum\limits_{n_{{}_3}=0}^{+\infty} p\left(n_{{}_3}|T\right)
\delta\left[T_\mathrm{meas}-
T_\mathrm{meas}\!\left(n_{{}_3}\right)\right].
\label{cond_PDTC_Poisson}
\end{equation}
From this distribution the conditional expectation value $\E$ of
$T_\mathrm{meas}^2$ is found to be equal to $T^2$,
\begin{equation}
\E\!\left(T_\mathrm{meas}^2|T\right)=T^2.\label{CondMeanTr}
\end{equation}
The corresponding conditional variance reads as
\begin{equation}
\Var\! \left(T_\mathrm{meas}^2|T\right)\!=\frac{T^2}
{\eta_{{}_3}\left|T_2\right|^2r^2}
+\frac{\bar{N}_3}{\eta_{{}_3}^2\left|T_2\right|^4r^4}.
\label{CondVarTr}
\end{equation}
Equations~(\ref{CondMeanTr})~and~(\ref{CondVarTr}) yield the
relative error of $T^2$,
\begin{eqnarray}
\epsilon=&&\frac{\sqrt{\Var\! \left(T_\mathrm{meas}^2|T\right)}}
{\E\!\left(T_\mathrm{meas}^2|T\right)}\nonumber\\&&=
\sqrt{\frac{1}{\eta_{{}_3}\left|T_2\right|^2T^2r^2}
+\frac{\bar{N}_3}{\eta_{{}_3}^2\left|T_2\right|^4T^4r^4}}.\label{Error}
\end{eqnarray}
From this expression, it follows that for the measurement of $T^2$
with the relative error $\epsilon$ one has to use the
local-oscillator amplitude
\begin{equation}
r=\frac{1}{T\left|T_2\right|\sqrt{\eta_{{}_3}}\epsilon}
\sqrt{\frac{1}{2}+ \sqrt{\frac{1}{4}+
\epsilon^2\bar{N}_3}}.\label{LocalOscillator}
\end{equation}
Hence a problem of monitoring appears for tiny $T$ values, for which
the value of the local-oscillator amplitude should be really large.
However, for this domain the signal is of poor quality anyway. We
can choose a minimal value $T_\mathrm{min}$ of the transmission
coefficient $T$ and post-select the values $T\geq T_\mathrm{min}$.
Based on this assumption, Eq.~(\ref{cond_PDTC_Poisson}) with the
relative error $\epsilon$ for $T^2$ reduces to
\begin{equation}
\mathcal{P}\!\left(T_\mathrm{meas}|T\right)=\delta\left(T-T_\mathrm{meas}\right),
\label{cond_PDTC_delta}
\end{equation}
in which connection the local-oscillator amplitude should be chosen
according to Eq.~(\ref{LocalOscillator}) with $T{=}T_\mathrm{min}$.
Under these conditions the shot noise error of detector
$\mathrm{D}_3$ becomes small.

Utilizing the $\delta$-function form~(\ref{cond_PDTC_delta}) for the
conditional PDTC in Eq.~(\ref{P_Monit_Noisy}), one gets
\begin{eqnarray}
P_\mathrm{noisy}\!\left(\alpha\right)&&=\int\limits_{0}^{1}\D
T\mathcal{P}\!\left(T\right)\label{P_Monit}\\
&&\times\exp\left[\frac{\bar N_\mathrm{nc}} {4 |R_2|^2 r^2
T^2\eta}\Delta_\alpha\right]\frac{1}{T^2\eta}
P\!\left(\frac{\alpha}{T\sqrt{\eta}}\right).\nonumber
\end{eqnarray}
The effect of noise counts can be omitted if the local-oscillator
amplitude obeys the condition
\begin{equation}
r^2\gg\frac{\bar N_\mathrm{nc}}{|R_2|^2 T^2_\mathrm{min}\eta}.
\label{LocalOscillatorNoise}
\end{equation}
In this case, the input-output relation~(\ref{P_Monit}) reduces to
\begin{equation}
P_\mathrm{noisy}\!\left(\alpha\right)=\int\limits_{0}^{1}\D
T\mathcal{P}\!\left(T\right)\frac{1}{T^2\eta}
P\!\left(\frac{\alpha}{T\sqrt{\eta}}\right),\label{P_Monit_N0}
\end{equation}
which appears to be similar to the case of an independently
controlled local oscillator, (cf. Ref.~\cite{Semenov1}). However, an
important difference is that this relation does not contain phase
noise of the signal after passing through the turbulent channel. It
is also worth noting that the $\delta$-function form of the
conditional PDTC~(\ref{cond_PDTC_delta}) excludes the
postprocessing noise of the measured data. For this reason, no fake
quantum effects appear in
Eqs.~(\ref{P_Monit})~and~(\ref{P_Monit_N0}).

\begin{figure}[ht!]
\includegraphics{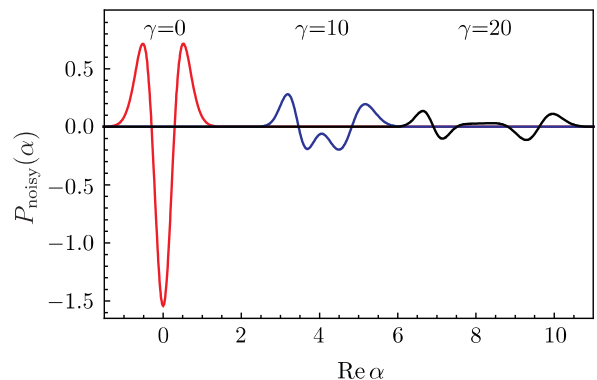}\\
\caption{\label{fig:SPATS_P} (Color online) $P$ function of the
displaced SPATS for $\Im\alpha{=}0$, with $n_\mathrm{th}=1.11$ (such
as realized in Ref.~\cite{Kiesel}) and different values of $\gamma$,
disturbed by beam wandering and detected with monitored transmission
coefficient. In the corresponding turbulence model
(cf.~the appendix~\ref{AppendixA}), the beam-spot radius, $W$, on the
receiver aperture of radius $a$ is $W{=}0.9 a$, and the standard
deviation of beam-deflection is $\sigma{=}10 a$. The detection
efficiency is $\eta{=}0.5$. }
\end{figure}

As in the previous section, we  illustrate the method with
the single-photon-added thermal states [cf. Eq.~(\ref{SPATS_0})].
Besides, we  suppose that this state is displaced with the
coherent amplitude $\gamma$ (cf.~Ref.~\cite{Semenov1}) such that its
$P$ function is
\begin{equation}
P\left(\alpha\right)=\frac{1}{\pi
\bar{n}_\mathrm{th}^3}\left[\left(1+\bar{n}_\mathrm{th}
\right)\left|\alpha-\gamma\right|^2-\bar{n}_\mathrm{th}
\right]e^{-\frac{\left|\alpha-\gamma\right|^2}{\bar{n}_\mathrm{th}}}.
\label{SPATS}
\end{equation}
As  has been shown in Ref.~\cite{Semenov1}, increasing the
displacement amplitude $\gamma$ results in diminishing the
nonclassicality in the scenario when the signal and local oscillator
are radiated from different sources. It turns out that this rule
does not apply  in the considered case. The absence of phase
fluctuations in input-output relation~(\ref{P_Monit_N0}) lifts
strong restrictions for the coherent amplitude of nonclassical
states considered in Ref.~\cite{Semenov1}. The quantum state in this
case may preserve its nonclassical properties. In
Fig.~\ref{fig:SPATS_P}, we show the $P$~function for the scenario with
the monitored transmission coefficient. It is clearly seen that
state is still nonclassical even for large values of the
coherent displacement amplitude $\gamma$.

\section{Quadrature squeezing}
\label{Quadrature}

Quadrature squeezing is a remarkable property of quantum light,
which can be observed by homodyne detection. In this section,
we consider how the disturbance of the signal and the local oscillator by
the atmosphere affects the detection of quadrature squeezing. We
consider two orthogonal quadratures [cf. Eq.~(\ref{Eq:Quadrature})], for a
certain value of the local-oscillator phase $\varphi$,
\begin{eqnarray}
&&\hat x_1=\hat x\left(\varphi\right),\label{SquezzedQ}\\
&&\hat x_2=\hat
x\left(\varphi+\frac{\pi}{2}\right).\label{AntiSquezzedQ}
\end{eqnarray}
A well known relation,
\begin{equation}
\left\langle \Delta \hat x_i \Delta \hat x_j\right\rangle=
\frac{1}{2}\delta_{i,j}+\left\langle :\Delta\hat x_i \Delta\hat
x_j:\right\rangle, \label{CovGen}
\end{equation}
$i,j{=}1,\,2$, connects the covariance matrix with its
normally ordered form. If a diagonal element (variance) of the
latter becomes negative for properly chosen $\varphi$, the state is
quadrature squeezed.

Let us first consider the case with monitored transmission
coefficient; see Sec.~\ref{Monitored}. Based on Eq.~(\ref{P_Monit}),
we can write the input-output relation for the covariance matrix,
\begin{eqnarray}
&&\left\langle :\Delta\hat x_i \Delta\hat
x_j:\right\rangle_\mathrm{noisy}=\label{CovMon}\\
&&\eta\left\langle T^2\right\rangle\left\langle :\Delta\hat x_i
\Delta\hat x_j:\right\rangle +\eta\left\langle\hat x_i
\right\rangle\left\langle\hat x_j \right\rangle\left\langle \Delta
 T^2\right\rangle \nonumber\\
&& +\frac{\bar N_\mathrm{nc}\left\langle T^{-2}\right\rangle}{\eta
r^2 \left|R_2\right|^2}\delta_{i,j}. \nonumber
\end{eqnarray}
The first term of this relation resembles the standard
attenuation. The second term is caused by the atmospheric
turbulence. The third term of the equation describes the disturbance
effect of noise counts on the quadrature squeezing.

The contribution from noise counts disappears for a sufficiently strong
local oscillator when $r^2{\gg} \frac{\bar N_\mathrm{nc}\left\langle
T^{-2}\right\rangle}{\eta \left|R_2\right|^2}$. However, due to
large contributions of events with small $T$, the value of
$\left\langle T^{-2}\right\rangle$ can be really large. For example,
if fluctuating losses are caused by beam wandering (cf.
the appendix~\ref{AppendixA} and Ref.~\cite{Vasylyev}), this term is
infinite. In practice, however, the measured transmission
coefficient is bounded by its minimal value $T_\mathrm{min}$. This
implies that the third term of Eq.~(\ref{CovMon}) becomes negligible
 for the local-oscillator amplitude satisfying
condition~(\ref{LocalOscillatorNoise}).

It is readily seen from the second term of Eq.~(\ref{CovMon}) that
the disturbance effect of the turbulence on the quadrature squeezing
increases with increasing mean value of the quadrature
$\left\langle\hat x_i \right\rangle$. This means that states with
a small coherent amplitude have better chances of preserving this
nonclassical property. For the states with $\left\langle\hat x_i
\right\rangle{=}0$, the squeezing of the $i\mathrm{th}$ quadrature
is disturbed in the same way as for standard attenuation.

Next we consider the case of non-monitored transmission coefficient;
see Sec.~\ref{Effective}. The corresponding input-output relation
for the covariance matrix is obtained from Eq.~(\ref{P_NonMonit})
and reads as
\begin{eqnarray}
&&\left\langle :\Delta\hat x_i \Delta\hat
x_j:\right\rangle_\mathrm{noisy}=\label{CovNonMon}\\
&&\eta\left\langle T_\mathrm{eff}^2\right\rangle\left\langle
:\Delta\hat x_i \Delta\hat x_j:\right\rangle +\eta\left\langle\hat
x_i \right\rangle\left\langle\hat x_j \right\rangle\left\langle
\Delta
T_\mathrm{eff}^2\right\rangle\nonumber\\
&&+\left(\frac{\left\langle
T_\mathrm{eff}\right\rangle-1}{2}+\frac{\bar N_\mathrm{nc}}{\eta r^2
T_\mathrm{ref}^{2}}\right)\delta_{i,j}, \nonumber
\end{eqnarray}
where $T_\mathrm{eff}$ is given by Eq.~(\ref{TrCoeffEff}). In contrast
to the case of monitored transmission coefficient, the third term in
this equation does not disappear even in the case of weak noise
counts.

\begin{figure}[ht!]
\includegraphics{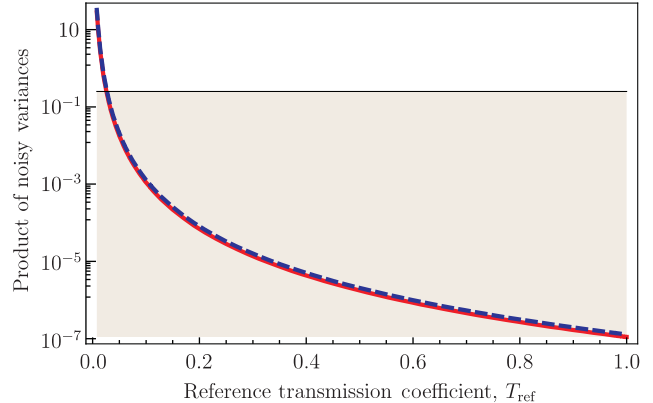}
\caption{\label{fig:UncertaintyRelations} (Color online) Product of
noisy variances, $\left\langle\Delta \hat
x_{1}^{2}\right\rangle_\mathrm{noisy}\left\langle\Delta \hat
x_{2}^{2}\right\rangle_\mathrm{noisy}$, vs the reference
transmission coefficient, $T_\mathrm{ref}$, for the scheme with
non-monitored turbulence. The solid (dashed) line corresponds to the
vacuum state ($8$-$\mathrm{dB}$ squeezed vacuum state) at the
transmitter. We suppose that fluctuating losses are caused by beam
wandering with the standard deviation of beam deflection
$\sigma{=}40a$ and with the beam-spot radius $W{=}0.95 a$, which
leads to $35$-$\mathrm{dB}$ mean losses. The detection efficiency is
$\eta{=}0.5$. The shaded area corresponds to the violation of the
Heisenberg uncertainty relation.}
\end{figure}

The post-processing noise may result in such an effective covariance
matrix that the corresponding density operator is not positive
semidefinite anymore. This means that the Heisenberg uncertainty
relation,
\begin{equation}
\left\langle\Delta \hat
x_{1}^{2}\right\rangle_\mathrm{noisy}\left\langle\Delta \hat
x_{2}^{2}\right\rangle_\mathrm{noisy}\geq \frac{1}{4},
\label{eq:UncertaintyRelations}
\end{equation}
is not satisfied; see Fig.~\ref{fig:UncertaintyRelations}. When the
real transmission coefficient appears to be much less compared with
the reference transmission coefficient $T_\mathrm{ref}$, the
negative effective noise [cf.~Eq.~(\ref{NcountsEff})] leads to such
a nonphysical squeezing.

The covariance matrix completely characterizes quantum states in the
Gaussian approximation. They can be used, for example, for
continuous-variable protocols of quantum-key distribution with
coherent and squeezed states~\cite{Scarani}. Evidently, in order to
get a consistent effective density operator in this approximation,
one should use an appropriate reference transmission coefficient
$T_\mathrm{ref}$. This means that the effect of the post-processing
noise may preserve consistency of the noisy density operator even in
the case of using the scheme with non-monitored transmission
coefficient.

\section{Summary and conclusions}
\label{Conclusions}

Homodyne detection of quantum light passing through the turbulent
atmosphere with the local oscillator sent in the
orthogonally polarized mode is a promising technique for
long-distance quantum communication based on continuous variables.
In this connection, a problem appears of how to connect a measured
photocount difference with the field quadrature. Indeed, when the
receiving local-oscillator amplitude is a fluctuating variable, this
question is not trivial anymore. We consider two possible solutions
for this problem. One possibility could be based on some reference
value for the local-oscillator amplitude transmission coefficient
(e.g., its mean value). Alternatively, here we proposed a method based
on the monitoring of the transmitted local oscillator with the aim of
 having control over the fluctuating transmission coefficient of the
turbulent atmosphere. In both cases, the quantum state of the
received light can be characterized by a noisy density operator,
which includes also information about shortcomings of the
measurement and postprocessing procedures of the used methods.

When the local-oscillator amplitude (and thus the transmission
coefficient) is monitored, the main limitations are caused by
stray-light and dark-count noise. These effects can be, in
principle, eliminated by a sufficiently strong local oscillator and
the postselection of events with an appropriately chosen threshold
value of the transmission coefficient. In the simpler procedure,
using a fixed reference value of the local-oscillator amplitude, the
shortcomings caused by the resulting postprocessing noise are much
more dramatic. The resulting disadvantages cannot be eliminated
anymore. In such a scenario, the noisy density operator may even
violate the fundamental requirement of positive semidefiniteness.
Thus fake quantum effects may occur due to the used post-processing
procedure. However, even based on this technique one may obtain a
consistent noisy density operator in the Gaussian approximation,
provided that the reference transmission coefficient is properly
chosen. We believe that these methods may be of some interest in the
context of continuous-variable quantum key distribution.

\begin{acknowledgements}
The authors gratefully acknowledge useful discussions with Ch.
Marquardt, B. Heim, and V. C. Usenko.
\end{acknowledgements}

\appendix
\section{PDTC for beam wandering}
\label{AppendixA}

Here we remind readers of some results of Ref.~\cite{Vasylyev}, in particular
the explicit form of the PDTC when the fluctuating losses are caused
by beam wandering. If the beam is randomly deflected around the
aperture center according to a two-dimensional Gaussian distribution
with the variance $\sigma^2$, the PDTC is given by the log-negative
Weibull distribution,
\begin{equation}
\begin{split}
\mathcal{P}(T){=}&\frac{2 R^2}{\sigma^2 \lambda T} \Bigl(2\ln
\frac{T_0}{T}\Bigr)^{\frac{2}{\lambda}{-}1}\exp\Bigl[-\frac{1}{2\sigma^2}R^2\Bigl(2\ln
\frac{T_0}{T}\Bigr)^\frac{2}{\lambda}\Bigr]\label{Weibull}
\end{split}
\end{equation}
for $T\in \left[0, T_0\right]$ and $0$ else. Here the parameters
$T_0$, $\lambda$, and $R$ are expressed in terms of the beam-spot
radius at the aperture $W$ and the aperture radius $a$,
\begin{equation}
T_0 =\sqrt{1-\exp\Bigl[-2\frac{a^2}{W^2}\Bigr]},
\end{equation}
\vspace{-4ex}
\begin{equation}
\begin{split}
\lambda =&8\frac{a^2}{W^2} \frac{\exp\bigl[-4\frac{a^2}{W^2}\bigr] {
\I}_1\bigl(4\frac{a^2}{W^2}\bigr)}{1-\exp[-4 \frac{a^2}{W^2}]{
\I}_0\bigl(4\frac{a^2}{W^2}\bigr)}\\
&\times\Bigl[\ln\Bigl(\frac{2T_0^2}{1-\exp[-4 \frac{a^2}{W^2}] {
\I}_0\bigl(4\frac{a^2}{W^2}\bigr)}\Bigr)\Bigr]^{-1}
\end{split}
\end{equation}
\vspace{-2ex}
\begin{equation}
\label{lambda}
 R=a \Bigl[\ln\Bigl(\frac{2T_0^2}{1-\exp[-4 \frac{a^2}{W^2}]
{\I}_0\bigl(4\frac{a^2}{W^2}\bigr)}\Bigr)\Bigr]^{-\frac{1}{\lambda}}.
\end{equation}
In the case when the turbulence is weak and the beam is focused on
the aperture, the beam-deflection variance can be approximately
evaluated as
\begin{equation}
\sigma^2\approx 1.919\, C_n^2 z^3 (2 W_0)^{-1/3},
\end{equation}
where $C_n^2$ is the index-of-refraction structure constant, $W_0$
is the beam-spot radius at the radiation source, and $z$ is the distance
between source and receiver aperture~\cite{Fante, Berman}.
Integration with the PDTC $\mathcal{P}(T)$ must be performed in the
Lebesgue sense with respect to the measure $\D
\left[R(2\ln\frac{T_0}{T})^\frac{1}{\lambda}\right]$.

\end{document}